\documentclass[a4paper,11pt]{article}
\pdfoutput=1
\usepackage{jcappub}
\usepackage[T1]{fontenc}
\usepackage{float}

\title{\boldmath Hunting Thermal Relics in the DESI DR1 Ly$\alpha$ Forest}

\author[a,1]{Emanuelly Silva \note{Corresponding author.}}
\author[a]{Artur Ladeira}
\author[a,b]{Rafael C. Nunes}
\author[c]{Eleonora Di Valentino}

\affiliation[a]{Instituto de Física, Universidade Federal do Rio Grande do Sul, 91501-970 Porto Alegre RS, Brazil}
\affiliation[b]{Divisão de Astrofísica, Instituto Nacional de Pesquisas Espaciais, Avenida dos Astronautas 1758, São José dos Campos, 12227-010, São Paulo, Brazil}
\affiliation[c]{School of Mathematical and Physical Sciences, University of Sheffield, UK}

\emailAdd{emanuelly.santos@ufrgs.br}
\emailAdd{artur.ladeira@ufrgs.br}
\emailAdd{rafadcnunes@gmail.com}
\emailAdd{e.divalentino@sheffield.ac.uk}

\abstract{We investigate constraints on additional relativistic species and thermal sterile neutrinos using the DESI DR1 one-dimensional Lyman-$\alpha$ forest power spectrum, combined with Planck 2018 CMB and DESI DR2 BAO measurements. We consider both the $\Lambda$CDM+$N_{\rm eff}$ extension and a thermal with a different-temperature sterile-neutrino (DTS) scenario, in which the sterile relic can be colder than the standard neutrino background. We first validate the DESI two-parameter $P_{\rm 1D}$ compression for the DTS model, finding that the residual cosmological dependence not captured by the compressed parameters remains below $0.15\%$. No significant evidence for additional radiation or a sterile component is found. For $\Lambda$CDM+$N_{\rm eff}$, we obtain $N_{\rm eff}<3.41$ at 95\% credibility from CMB+DESI-BAO+DESI-$P_{\rm 1D}$. In the DTS scenario, the full CMB+DESI-BAO+DESI-$P_{\rm 1D}$ combination yields the stringent bound $m_s^{\rm eff}<0.061\,{\rm eV}$, highlighting the complementarity of BAO and Lyman-$\alpha$ information in constraining the massive sterile abundance. We further interpret the allowed $\Delta N_{\rm eff}$ in terms of thermal light relics, deriving lower limits on their decoupling temperatures that reach the QCD epoch.}

\begin{document}
\maketitle
\flushbottom

\section{Introduction}
\label{sec:intro}

The effective number of relativistic species, $N_{\rm eff}$, provides a sensitive probe of the radiation content of the early Universe and links cosmology to particle physics~\cite{Cielo:2023bqp,DiValentino:2024xsv,Escudero:2024uea}. In the Standard Model, non-instantaneous neutrino decoupling, flavor oscillations, and finite-temperature QED corrections give the precise prediction $N_{\rm eff}^{\rm SM}\simeq 3.044$~\cite{Akita:2020szl,Bennett:2020zkv,Drewes:2024wbw}. Any deviation from this value could indicate a nonstandard thermal history or additional light degrees of freedom in the primordial plasma. In particular, $\Delta N_{\rm eff}>0$ can result from light thermal relics, including scalar or axion-like particles, hidden-sector fermions, dark gauge bosons, and sterile neutrinos~\cite{Brust:2013ova,Weinberg:2013kea,Giusarma:2014zza,DiValentino:2015zta,CMB-S4:2016ple,Baumann:2017gkg,Giare:2020vzo,Carenza:2021ebx,Giare:2021cqr,DEramo:2022nvb,Dvorkin:2022jyg,Caloni:2024olo}. Conversely, an inferred $\Delta N_{\rm eff}<0$ may arise in scenarios that preferentially heat the electromagnetic plasma relative to the neutrino sector after neutrino decoupling, as can occur for MeV-scale particles coupled to electrons and photons~\cite{Serpico:2004nm,Ho:2012ug,Boehm:2013jpa,Nollett:2013pwa,Escudero:2018mvt,Sabti:2019mhn}. Cosmologically, $\Delta N_{\rm eff}$ is also relevant to the Hubble tension~\cite{Verde:2019ivm,DiValentino:2020zio,DiValentino:2021izs,Perivolaropoulos:2021jda,Abdalla:2022yfr,DiValentino:2022fjm,Kamionkowski:2022pkx,Hu:2023jqc,Verde:2023lmm,DiValentino:2024yew,Cai:2026swf}, the discrepancy between early- and late-Universe measurements of $H_0$, which currently exceeds $7\sigma$~\cite{H0DN:2025lyy}. A positive $\Delta N_{\rm eff}$ increases the early-Universe expansion rate, reduces the sound horizon, and can raise the inferred value of $H_0$, potentially alleviating the tension (see Ref.~\cite{CosmoVerse:2025txj} for a recent review).

Among possible neutrino-sector extensions, light sterile neutrinos remain among the most studied (see Refs.~\cite{Gariazzo:2015rra,Abazajian:2017tcc,Dasgupta:2021ies,Acero:2022wqg} for reviews). The LSND experiment reported a $3.8\sigma$ excess of $\bar{\nu}_e$ events~\cite{LSND:2001aii}, while MiniBooNE observed a $4.8\sigma$ excess of low-energy electron-like events relative to the expected background~\cite{MiniBooNE:2020pnu}. These anomalies have motivated systematic investigations of additional neutrino states with masses around the eV scale~\cite{Kopp:2013vaa,Gariazzo:2017fdh,Dentler:2018sju,Hu:2025lrl,GarciaEscudero:2025orc,Goswami:2026qpl}. More recent short-baseline searches, however, have placed significant pressure on the simplest light-sterile interpretation~\cite{MicroBooNE:2025nll,Rodrigues:2025tha}.

Cosmological observables provide an independent and highly sensitive probe of light sterile neutrinos through their contributions to the radiation density, the expansion history, and the growth of matter perturbations~\cite{Planck:2018vyg,Archidiacono:2022ich,Du:2025iow,Ladeira:2026pck}. Current cosmological data strongly constrain massive sterile neutrinos when they are assumed to be fully thermalized. These limits, however, depend on the production mechanism and phase-space distribution of the sterile population and therefore do not generically exclude all sterile-neutrino realizations~\cite{Abazajian:2017tcc,Archidiacono:2022ich,Ladeira:2026pck}.

A physically motivated alternative to the standard sterile-neutrino scenario is the thermal sterile-neutrino framework with a different temperature (DTS). Here, the sterile species follows a thermal momentum distribution with a temperature $T_s$ that differs from that of the active-neutrino background, $T_\nu$~\cite{Feng:2008ya,Das:2010ts,Ladeira:2026pck}. Such a thermal history may arise if the sterile sector decouples differently from the Standard Model plasma and subsequently evolves with its own temperature. For $T_s<T_\nu$, the sterile population is colder and less abundant than a fully thermalized species. Its contribution to both the relativistic energy density at early times and the hot-dark-matter abundance at late times is consequently suppressed. A comparatively large physical sterile mass can therefore correspond to a substantially smaller cosmological energy density than in the fully thermalized limit~\cite{Acero:2008rh,Das:2021pof,Ladeira:2026pck}.

Assessing the viability of these and other scenarios requires increasingly precise constraints on their parameter spaces. Fortunately, from a cosmological perspective, we are in a golden era of data, enabling a wide range of tests and increasingly accurate measurements. Recently, Refs.~\cite{Goldstein:2026iuu,Loverde:2026cxe,Akharman:2026izy} have combined state-of-the-art measurements of the primordial helium and deuterium abundances with Cosmic Microwave Background (CMB) and Baryon Aacoustic Oscillations (BAO) data to derive the tightest constraints on $N_{\rm eff}$ to date (see also~\cite{Yeh:2022heq,Ganguly:2025mdi,Escudero:2026mgw}). Recent full-shape power-spectrum measurements of galaxies have also helped break parameter degeneracies and provide improved sensitivity to the neutrino mass and non-standard properties~\cite{Kumar:2022vee,Elbers:2025vlz,Chudaykin:2025lww,Noriega:2025ulc,Das:2025asx,Poudou:2025qcx,Ivanov:2026dvl} for results along similar lines. Furthermore, scenarios with new background dynamics involving neutrinos are also being explored using these cutting-edge data~\cite{Reboucas:2024smm,Yadav:2024duq,Jiang:2024viw,Du:2024pai,Du:2025xes,Sharma:2025iux,Sabogal:2025qhz,Anchordoqui:2025elg,Barua:2025adv,Feng:2025mlo,Du:2025iow,Ladeira:2026jne,Pulido-Hernandez:2026hcs,Yang:2026yaq,Li:2026ldf,Li:2026asg,Zhou:2026dti,Qu:2026fkt,LSSTDarkEnergyScience:2026ach,Giare:2026oti,Kibris:2026cqq,Feng:2026pzs,Montandon:2026vuc}.

More recently, the DESI Collaboration released the first compressed parameters for the one-dimensional Lyman-$\alpha$ forest flux power spectrum~\cite{Chaves-Montero:2026hqd}. These parameters retain most of the cosmological information on additional relativistic species and are directly related to the linear matter power spectrum. The latter carries several characteristic signatures of neutrinos. Massive neutrinos suppress power on small scales through free streaming, with the effect generally increasing with $\sum m_\nu$. Additional relativistic species can also modify the overall shape of the spectrum. In particular, a $\Delta N_{\rm eff} \neq 0$ can shift matter-radiation equality and consequently alter the subsequent growth of structure. Finally, both $\sum m_\nu$ and $N_{\rm eff}$ can affect the BAO observables. Together, these effects make the Lyman-$\alpha$ forest a sensitive probe of neutrino masses and additional relativistic degrees of freedom.

\textit{In this work, we use the DESI DR1 compressed 1D Lyman-$\alpha$ flux power spectrum to derive new constraints on the $\Lambda$CDM+$N_{\rm eff}$ and DTS models. The $\Lambda$CDM+$N_{\rm eff}$ case has already been investigated in Ref.~\cite{Chaves-Montero:2026hqd}, while here we extend the analysis to the DTS model and explore the thermal-relic interpretation of the constraints on additional relativistic species within the $\Lambda$CDM+$N_{\rm eff}$ framework.} The \textit{paper} is organized as follows. Section~\ref{sec:neutrino_models} introduces the $\Lambda$CDM+$N_{\rm eff}$ and DTS scenarios. Section~\ref{sec:methodology} outlines the methodology, including the datasets, analysis setup, and statistical tests used to compare the models. Our main results are presented in Section~\ref{sec:results}, where we first assess the validity of the compressed Lyman-$\alpha$ description for the DTS scenario and subsequently present the cosmological constraints on both models and the thermal-relic interpretation of the inferred $\Delta N_{\rm eff}$ constraints. Section~\ref{sec:conclusions} summarizes our main conclusions.

\section{Neutrino-sector models}
\label{sec:neutrino_models}

In this section, we briefly introduce the two extensions of the standard neutrino sector considered in this work. The first is the $\Lambda$CDM+$N_{\rm eff}$ model, in which the effective number of relativistic species is allowed to vary. The second is the DTS scenario~\cite{Ladeira:2026pck}, which introduces a massive thermal sterile neutrino with a temperature different from that of the active-neutrino background. We describe both scenarios below.

\subsection{Effective number of relativistic species}
\label{sec:neff}

The total radiation energy density after electron-positron annihilation is conventionally parameterized in terms of the effective number of relativistic species, $N_{\rm eff}$, as

\begin{equation}
    \rho_{\rm rad}
    =
    \rho_{\gamma}
    \left[
        1+
        \frac{7}{8}
        \left(\frac{4}{11}\right)^{4/3}
        N_{\rm eff}
    \right],
    \label{eq:rho_rad_neff}
\end{equation}
where $\rho_{\gamma}$ denotes the photon energy density. Although expressed in neutrino units, $N_{\rm eff}$ should not be interpreted as a literal count of neutrino species. Rather, it serves as a phenomenological measure of the total relativistic energy density beyond photons and may thus encode the presence of additional light degrees of freedom in the early Universe.

As previously mentioned, non-instantaneous neutrino decoupling, flavor oscillations, and finite-temperature QED corrections yield the Standard Model prediction $N_{\rm eff}^{\rm SM}=3.044$~\cite{Akita:2020szl,Bennett:2020zkv,Drewes:2024wbw}. Deviations from this value can be parameterized as
\begin{equation}
    \Delta N_{\rm eff}
    \equiv
    N_{\rm eff}-N_{\rm eff}^{\rm SM}.
    \label{eq:delta_neff}
\end{equation}

Here, we treat $N_{\rm eff}$ as a free cosmological parameter, remaining agnostic about the microscopic origin of a possible departure. We thus interpret $\Delta N_{\rm eff}$ solely as a modification of the early-Universe radiation density, without specifying the underlying particle content or production mechanism. Any additional relativistic component is assumed to behave as collisionless, free-streaming radiation, following the conventional $\Lambda$CDM+$N_{\rm eff}$ extension~\cite{Planck:2018vyg,AtacamaCosmologyTelescope:2025nti}.

Changes in $N_{\rm eff}$ alter the expansion history during radiation domination and affect the matter power spectrum by modifying the transfer of primordial fluctuations. Because our Lyman-$\alpha$ likelihood is based directly on the power spectrum $P(k)$, it provides a sensitive, complementary probe of these effects, alongside CMB and BAO measurements, which constrain the expansion history as well as, in the case of the CMB, the evolution of perturbations. In Sec.~\ref{sec:thermal_interpretation}, we interpret the resulting constraints on $\Delta N_{\rm eff}$ in terms of possible particle-physics scenarios.

\subsection{Thermal with a different-temperature sterile-neutrino (DTS)}
\label{sec:dts}

We next consider a thermal sterile-neutrino relic whose temperature, $T_s$, is allowed to differ from that of the standard active-neutrino background, $T_\nu$. Following Ref.~\cite{Ladeira:2026pck}, we refer to this realization as the DTS scenario.

The sterile species is assumed to follow a Fermi-Dirac distribution,
\begin{equation}
    f_s(p) = \left[ \exp\left(\frac{p}{T_s}\right) + 1 \right]^{-1},
    \label{eq:dts_distribution}
\end{equation}\smallskip
where $p$ denotes the physical momentum. As the energy density of a relativistic thermal relic scales with the fourth power of its temperature, its contribution to the effective number of relativistic species is given by
\begin{equation}
    \Delta N_{\rm eff} = \left(\frac{T_s}{T_\nu}\right)^4 .
    \label{eq:dts_delta_neff}
\end{equation}

Once the sterile species becomes non-relativistic, its energy density is determined by its mass and number density. For the thermal Fermi-Dirac distribution in~\eqref{eq:dts_distribution}, the number density scales as $T_s^3$. Relative to a standard thermal neutrino species, one therefore has $n_s/n_\nu=(T_s/T_\nu)^3$, and the present-day physical density can be written as
\begin{equation}
    \omega_s
    \equiv
    \Omega_s h^2
    =
    \frac{m_s}{94.05\,{\rm eV}}
    \left(\frac{T_s}{T_\nu}\right)^3
    =
    \frac{m_s^{\rm eff}}{94.05\,{\rm eV}},
    \label{eq:dts_density}
\end{equation}
where $m_s$ denotes the physical sterile-neutrino mass and $m_s^{\rm eff}$ is the effective mass parameter that determines its cosmological abundance. Combining Eqs.~\eqref{eq:dts_delta_neff} and~\eqref{eq:dts_density} gives
\begin{equation}
    m_s^{\rm eff}
    =
    m_s\left(\Delta N_{\rm eff}\right)^{3/4}.
    \label{eq:dts_meff}
\end{equation}

These relations show how the DTS scenario differs from a fully thermalized sterile species. When $T_s<T_\nu$, the sterile relic is colder and less abundant, so a given physical mass contributes less to both the early-time radiation density and the late-time hot-dark-matter abundance. The fully thermalized case is recovered when $T_s=T_\nu$, for which $\Delta N_{\rm eff}=1$ and $m_s^{\rm eff}=m_s$. Lowering the temperature ratio suppresses both quantities, allowing a relatively large physical sterile-neutrino mass to have a much smaller cosmological energy density~\cite{Acero:2008rh,Das:2021pof,Ladeira:2026pck}.

For the numerical implementation of the DTS scenario, we sample the additional relativistic contribution $\Delta N_{\rm eff}$ together with the physical sterile-neutrino mass $m_s$. At each MCMC step, $\Delta N_{\rm eff}$ is mapped onto the sterile-neutrino temperature according to
\begin{equation}
    T_s=T_\nu(\Delta N_{\rm eff})^{1/4},
\end{equation}
and the corresponding temperature ratio is passed to CLASS for the additional thermal \texttt{ncdm} species (we provide further details in Sec.~\ref{sec:analysis}).

\section{Methodology}
\label{sec:methodology}

This section is organized into three parts. First, we describe the observational datasets used to constrain the parameters of the cosmological models under consideration, with particular emphasis on the construction and implementation of the Lyman-$\alpha$ likelihood. We then outline the analysis setup, including the main stages of the computational pipeline, the adopted priors, and the convergence criteria. Finally, we briefly describe the statistical tools used to assess and compare the cosmological models.

\subsection{Datasets}

We constrain the cosmological models using the compressed constraints from the DESI DR1 1D Lyman-$\alpha$ flux power spectrum. We combine them with Planck CMB temperature and polarization data (TT, TE, and EE), CMB lensing measurements, and DESI DR2 BAO measurements. The datasets are described below.

\paragraph{1D Lyman-$\alpha$ power spectrum (\texttt{DESI-P$_{\rm 1D}$}).}
As shown in Ref.~\cite{Pedersen:2022anu}, over $2<z<5$, most of the cosmological information encoded in the one-dimensional Lyman-$\alpha$ forest flux power spectrum can be captured through a two-parameter compression. These parameters correspond to the amplitude and logarithmic slope of the linear matter power spectrum, evaluated at a pivot $k$ and $z$. They are defined as
\begin{equation}
\Delta_\star^2 \equiv
\frac{k_\star^3\,P_{\rm lin}(k_\star,z_\star)}{2\pi^2},
\qquad
n_\star \equiv
\left.\frac{d\ln P_{\rm lin}(k,z)}{d\ln k}\right|_{(k_\star,z_\star)}.
\end{equation}
For the DESI DR1 analysis, we adopt the pivot values $z_\star=3$ and $k_\star=0.009~\mathrm{s\,km^{-1}}$, following the convention established in Ref.~\cite{Chaves-Montero:2026hqd}.

This compression removes the need for explicit simulation interpolation during parameter inference. Instead, it provides a fast likelihood that can be evaluated efficiently within the MCMC framework~\cite{Pedersen:2022anu}. Following the methodology of Ref.~\cite{Goldstein:2023gnw}, the resulting log-likelihood is given by
\begin{align}
\log\mathcal{L} = -\frac{1}{2(1-\rho^2)}
\Big[ \Delta x^2 - 2\rho\,\Delta x\,\Delta y + \Delta y^2 \Big],
\label{eq:lyalpha_compressed_likelihood}
\end{align}
where
\begin{equation}
\Delta x \equiv
\frac{\Delta_\star^2 - \bar{\Delta}_\star^2}
{\sigma_{\Delta_\star^2}},
\qquad
\Delta y \equiv
\frac{n_\star - \bar{n}_\star}
{\sigma_{n_\star}}.
\end{equation}
The quantities $\bar{\Delta}_\star^2$ and $\bar{n}_\star$ denote the central values of the compressed DESI DR1 Lyman-$\alpha$ measurements, with $\sigma_{\Delta_\star^2}$ and $\sigma_{n_\star}$ their corresponding uncertainties. By contrast, $\Delta_\star^2$ and $n_\star$ are the theoretical predictions for each cosmological model at the chosen pivot scale and redshift. Finally, $\rho$ is the correlation coefficient between the two measurements and accounts for their covariance in the likelihood.

Our fiducial dataset is based on measurements of the one-dimensional Lyman-$\alpha$ forest flux power spectrum from DESI DR1~\cite{Karacayli:2025svi,Ravoux:2025uik}. It contains approximately $450000$ Ly$\alpha$ forests from the first year of the main survey and from Survey Validation observations~\cite{DESI:2023dwi}. The measurements were analyzed in Ref.~\cite{Chaves-Montero:2026hqd} using a hydrodynamical-simulation-based emulator that accounts for cosmological and IGM parameters, astrophysical contaminants, and observational systematics.

Throughout this work, we adopt the compressed Lyman-$\alpha$ measurements obtained from this analysis and refer to this dataset as \texttt{DESI-P$_{\rm 1D}$}.\footnote{Our likelihood is publicly available at \url{https://github.com/emanusilvas/Lyman-alpha-1D-likelihood}.} \bigskip

\noindent The additional datasets considered are:

\paragraph{Baryon acoustic oscillations (\texttt{DESI-BAO}).}
This dataset comprises baryon acoustic oscillation (BAO) measurements from the second DESI data release (DESI DR2). As reported in Table IV of Ref.~\cite{DESI:2025zgx}, the measurements are divided into seven effective redshift bins over the range $0.295 \leq z \leq 2.330$. The primary observables are the transverse comoving distance, $D_{\mathrm{M}}/r_{\mathrm{d}}$, the Hubble distance, $D_{\mathrm{H}}/r_{\mathrm{d}}$, and the volume-averaged distance, $D_{\mathrm{V}}/r_{\mathrm{d}}$, all normalized by the sound horizon at the drag epoch, $r_{\mathrm{d}}$. We account for the statistical correlations among these observables using the reported correlation coefficients, $r_{\mathrm{V,M/H}}$ and $r_{\mathrm{M,H}}$. We refer to this dataset as \texttt{DESI-BAO} throughout this work.

\paragraph{Cosmic microwave background (\texttt{CMB}).}
This dataset refers to the Planck 2018 legacy data~\cite{Planck:2018vyg}, including the temperature (TT), polarization (EE), and temperature-polarization cross-correlation (TE) power spectra. For high multipoles, we adopt the \texttt{Plik} likelihood for TT ($30 \leq \ell \leq 2508$) and for TE and EE ($30 \leq \ell \leq 1996$). At low multipoles, we use the \texttt{Commander} likelihood for TT and the \texttt{SimAll} likelihood for EE over $2 \leq \ell \leq 29$~\cite{Planck:2019nip}. We also include the Planck 2018 CMB lensing reconstruction~\cite{Planck:2018lbu}. We refer to the full combination of these datasets as \texttt{CMB}.

\subsection{Analysis setup}
\label{sec:analysis}

For the computation of the background cosmological evolution and linear perturbations for each cosmological model considered, we employ the Boltzmann code \texttt{CLASS}~\cite{Blas:2011rf}. We then use the publicly available code \texttt{MontePython}~\cite{Audren:2012wb,Brinckmann:2018cvx} to perform Markov Chain Monte Carlo (MCMC) sampling and carry out Bayesian parameter inference. To account for the specific sterile-neutrino scenario considered in this work, we modify the standard \texttt{MontePython} implementation to consistently derive the model-specific neutrino parameters of the DTS model throughout the MCMC analysis. For all chains, convergence is assessed using the Gelman-Rubin criterion~\cite{Gelman:1992zz}, requiring $R-1 < 0.02$.

Across all our analyses, we sample the physical baryon density $\omega_{\rm b} = \Omega_{\rm b}h^2$, the physical cold dark matter density $\omega_{\rm cdm} = \Omega_{\rm cdm}h^2$, the Hubble parameter today $H_0$, the amplitude of the primordial scalar power spectrum $\ln(10^{10}A_{\rm s})$, its spectral index $n_{\rm s}$, and the reionization optical depth $\tau_{\rm reio}$. For the neutrino sector, we consider two different scenarios. In the $\Lambda$CDM+$N_{\rm eff}$ model, we vary $N_{\rm eff}$ while fixing the properties of the massive-neutrino component to $T_\nu/T_\gamma=0.71611$ and $\sum m_{\nu}=0.06\,{\rm eV}$.

For the DTS scenario, we fix the standard neutrino sector contribution and add a massive thermal species. To avoid redundant parameters, we do not vary the sterile-neutrino temperature independently. Instead, we sample $\Delta N_{\rm eff}$ and the physical sterile-neutrino mass $m_s$, then derive the temperature from $T_s = T_\nu (\Delta N_{\rm eff})^{1/4}$ (see Sec.~\ref{sec:dts}). The effective mass, $m_s^{\rm eff}$, is likewise treated as a derived parameter in the posterior analysis. This parameterization retains the thermal phase-space distribution of the sterile relic while avoiding the introduction of a redundant parameter.

We adopt flat priors for all sampled parameters. For the standard cosmological sector, we consider
\begin{align*}
&\hspace{0.2cm}\omega_{\rm b} \in [0.020,0.025], \quad
\omega_{\rm cdm} \in [0.001,0.99], \quad
H_0 \in [20,100],\\
&\ln(10^{10}A_{\rm s}) \in [1.61,3.91], \quad
n_{\rm s} \in [0.8,1.2], \quad
\tau_{\rm reio} \in [0.004,0.8].
\end{align*}
For the neutrino sector, the corresponding prior ranges are
\begin{align*}
\Delta N_{\rm eff} \in [0,1.0], \qquad
m_s\,(\mathrm{eV}) \in [0,5.0].
\end{align*}

The compressed parameters introduced above, $\Delta_{\star}^{2}$ and $n_{\star}$, are derived from the linear matter power spectrum at each step of the MCMC analysis. Once the Markov chains were obtained, all post-processing was performed using the \texttt{GetDist}\footnote{Available at \url{https://github.com/cmbant/getdist}.} package~\cite{Lewis:2019xzd} to extract numerical results, such as one-dimensional posterior distributions and marginalized two-dimensional probability contours.

\subsection{Statistical tools for model comparison}
\label{sec:statistical_tools}

To compare the cosmological models considered in this work, we employ complementary statistical diagnostics. As a first measure, we consider the difference in the minimum $\chi^2$ values between each model and the reference $\Lambda$CDM model,
\begin{equation}
\Delta\chi^2_{\mathrm{min}} \equiv \chi^2_{\mathrm{min,\,MODEL}} - \chi^2_{\mathrm{min,\,\Lambda CDM}}\,.
\end{equation}
This provides a simple measure of the relative goodness of fit. A negative $\Delta\chi^2_{\rm min}$ indicates a better fit than $\Lambda$CDM, while a positive value indicates a worse fit. This statistic, however, does not account for differences in model complexity.

As a further check, we consider the Akaike Information Criterion (AIC)~\cite{Akaike:1974vps}, defined by
\begin{equation}
\mathrm{AIC} \equiv -2 \ln \mathcal{L}_{\mathrm{max}} + 2N,
\label{AIC}
\end{equation}
where $\mathcal{L}_{\mathrm{max}}$ denotes the maximum likelihood and $N$ is the total number of free parameters. Models with lower AIC values are preferred because the criterion weighs goodness of fit against model complexity. The $2N$ term penalizes additional parameters, helping to avoid unnecessary model extensions and overfitting.

Finally, to account for the full parameter space, we also consider the Bayes factor. Unlike criteria based on a single best-fit point, it incorporates the prior distributions and model complexity through the Bayesian evidence. For two models, $\mathcal{M}_i$ and $\mathcal{M}_j$, the Bayes factor is defined as
\begin{equation}
\mathcal{B}_{ij}
=
\frac{p(\mathbf{x}|\mathcal{M}_i)}
     {p(\mathbf{x}|\mathcal{M}_j)}
=
\frac{
\displaystyle \int d\boldsymbol{\theta}_i\,
\pi(\boldsymbol{\theta}_i)
\mathcal{L}(\mathbf{x}|\boldsymbol{\theta}_i)
}{
\displaystyle \int d\boldsymbol{\theta}_j\,
\pi(\boldsymbol{\theta}_j)
\mathcal{L}(\mathbf{x}|\boldsymbol{\theta}_j)
},
\label{BayesFactor}
\end{equation}
where $\pi(\boldsymbol{\theta})$ denotes the prior and $\mathcal{L}$ the likelihood. Assuming equal prior probabilities for the models, the Bayes factor is also equal to the ratio of their posterior probabilities. We evaluate $\ln\mathcal{B}_{ij}$ relative to $\Lambda$CDM, with $\Lambda$CDM in the numerator. Positive values therefore favor $\Lambda$CDM, while negative values favor the extended model. Following the scale proposed in Ref.~\cite{Kass:1995loi},
\begin{equation*}
|\ln \mathcal{B}_{ij}| =
\begin{cases}
< 1, & \text{Inconclusive}, \\
1-3, & \text{Positive evidence}, \\
3-5, & \text{Strong evidence}, \\
\geq 5, & \text{Very strong evidence}.
\end{cases}
\end{equation*}\smallskip

We compute the Bayes factors with the publicly available \texttt{MCEvidence}\footnote{Available at \url{https://github.com/yabebalFantaye/MCEvidence}.} package, which estimates the Bayesian evidence directly from posterior samples generated by the MCMC chains~\cite{Heavens:2017afc,Heavens:2017hkr}.

\section{Results}
\label{sec:results}

We divide our results into four subsections. In Subsec.~\ref{sec:compress}, we evaluate how well the compressed parameters retain the cosmological information of the models considered. Subsections~\ref{sec:Neff} and~\ref{sec:DTS} present the cosmological constraints for the two models analyzed. Finally, in Subsec.~\ref{sec:thermal_interpretation}, we briefly interpret these results in the context of particle physics.

\subsection{Efficiency of the compressed 1D Lyman-$\alpha$ parameters}
\label{sec:compress}

For the $\Lambda$CDM+$N_{\rm eff}$ model, Ref.~\cite{Chaves-Montero:2026hqd} has shown that the compressed parameters $\Delta_\star^2$ and $n_\star$ efficiently capture the relevant cosmological information. However, no such validation has been reported for the DTS model or for similar frameworks. We therefore assess the performance of this compression scheme for the DTS model following the methodology of Ref.~\cite{Chaves-Montero:2026hqd}.

We first compute the linear 1D power spectrum for the reference $\Lambda$CDM model using the Planck 2018 best-fit parameters,
\begin{equation}
\label{eq:p1d}
P_{\rm 1D}(k_\parallel) =
\frac{1}{2\pi}
\int_0^\infty \mathrm{d}k_\perp\, k_\perp\,
P_{\mathrm{lin}}(k_\parallel,k_\perp)
e^{-k^2/k_{\mathrm{pressure}}^2},
\end{equation}
where $k_\parallel$ and $k_\perp$ are the components of the wave vector $\mathbf{k}$ parallel and perpendicular to the line of sight, with $k^2=k_\parallel^2+k_\perp^2$ and $k_{\mathrm{pressure}}=0.4\,\mathrm{s/km}$. We then consider the DTS cosmology and assess the impact of its main parameters on $P_{\rm 1D}$. Finally, following Ref.~\cite{Pedersen:2022anu}, we rescale the primordial parameters $A_{\mathrm{s}}$ and $n_{\mathrm{s}}$ of the DTS cosmology. This procedure yields the corresponding rescaled $P_{\rm 1D}$ prediction. Any residual difference between the $P_{\rm 1D}$ values computed using Eq.~\ref{eq:p1d} for the rescaled DTS model and the baseline would indicate that the compressed parameters do not fully capture the observable's cosmological dependence.

\begin{figure}[H]
    \centering
    \includegraphics[width=\textwidth]{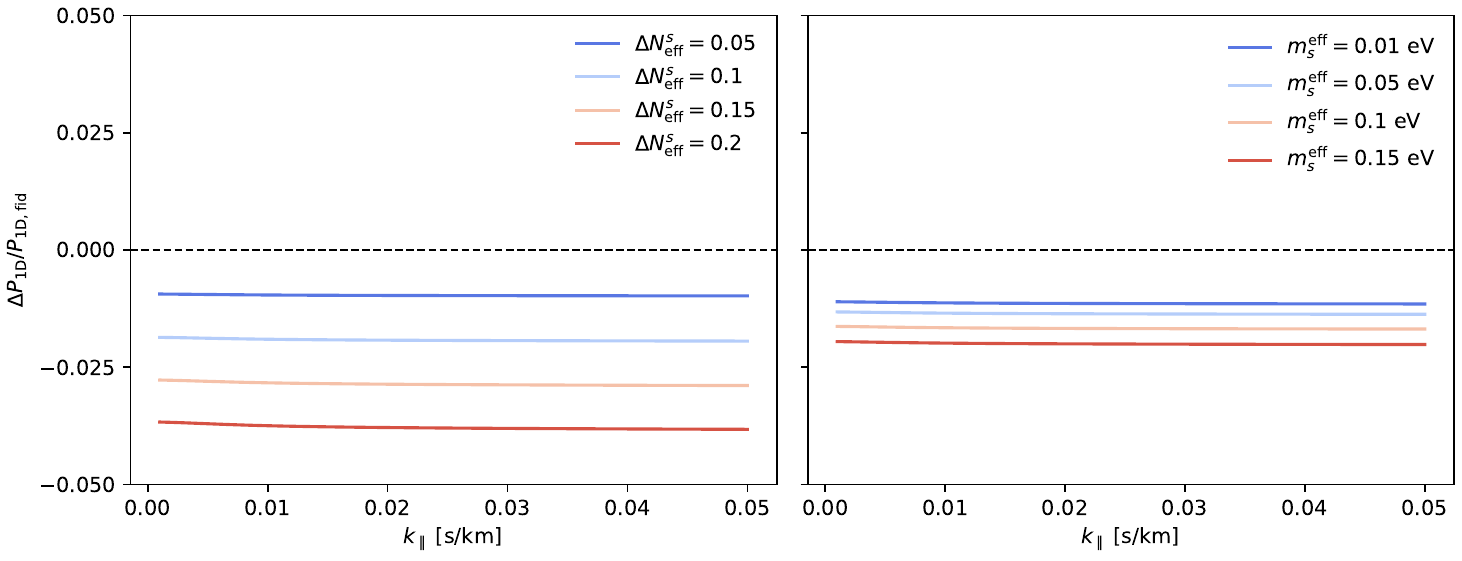}
    \caption{Residuals of the 1D linear power spectrum at $z=3$ between the DTS model, for various values of $\Delta N^{s}_{\mathrm{eff}}$ and $m^{\mathrm{eff}}_s$, and the fiducial $\Lambda$CDM cosmology.}
    \label{fig:Noscaling}
\end{figure}

As shown in Fig.~\ref{fig:Noscaling}, before rescaling, increasing
$\Delta N^{s}_{\mathrm{eff}}$ and/or $m^{\mathrm{eff}}_s$ in the DTS model
suppresses the power spectrum relative to $\Lambda$CDM, due to the impact
of the additional radiation density and sterile-neutrino free streaming
on the matter power spectrum. The resulting variations reach up to
$\sim 3\%$ over the range considered.

\begin{figure}[H]
    \centering
    \includegraphics[width=\textwidth]{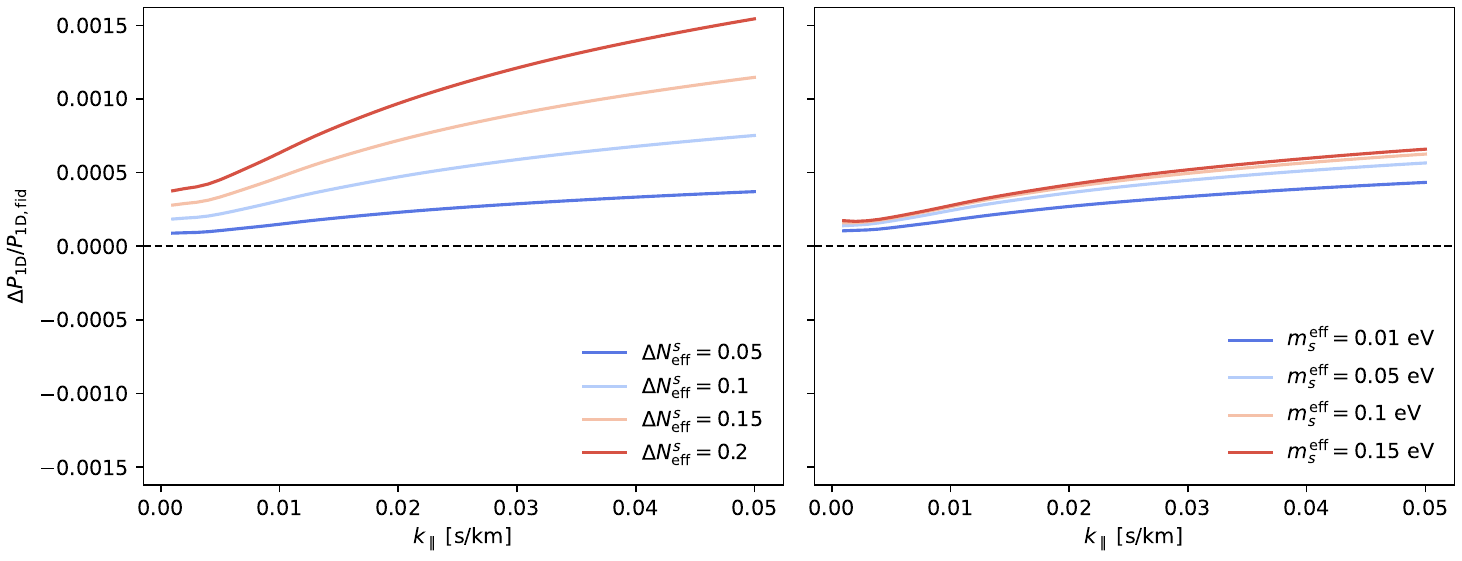}
    \caption{Same as Fig.~\ref{fig:Noscaling}, but after applying the rescaling procedure.}
    \label{fig:scaling}
\end{figure}

However, as shown in Fig.~\ref{fig:scaling}, these differences are strongly reduced after rescaling, with the maximum variation decreasing to $0.15\%$. This indicates that the compressed parameters $\Delta_\star^2$ and $n_\star$ effectively capture the relevant cosmological information of the DTS model over the scales considered.

Since these sub-percent residuals are well below the statistical uncertainties of current Lyman-$\alpha$ forest measurements, the two-parameter compression remains accurate at the redshift considered here, $z=3$. This result provides an initial validation of the application of the compressed likelihood to the DTS cosmology.

\subsection{Constraints on $\Lambda$CDM+$N_{\rm eff}$}
\label{sec:Neff}

Table~\ref{tab:Neffconstraints} summarizes our main results for this model. It reports the marginalized means and $68\%$ credible intervals for all explored parameters, except for $N_{\rm eff}$, for which we quote the $95\%$ upper limit. The resulting constraints on $N_{\rm eff}$ are consistently compatible with the Standard Model prediction, $N_{\rm eff}^{\rm SM}=3.044$. However, a positive contribution remains allowed at $95\%$ credibility:
\begin{equation*}
\Delta N_{\rm eff}
\begin{cases}
< 0.316, & {\rm CMB},\\
< 0.306, & {\rm CMB+DESI}\text{-}P_{\rm 1D},\\
< 0.396, & {\rm CMB+DESI\text{-}BAO},\\
< 0.366, & {\rm CMB+DESI\text{-}BAO+DESI}\text{-}P_{\rm 1D}.
\end{cases}
\label{eq:DNeff_limites}
\end{equation*}\smallskip

Possible physical interpretations of a positive $\Delta N_{\rm eff}$ are discussed in Subsec.~\ref{sec:thermal_interpretation}, particularly in connection with searches for light thermal relics~\cite{Baumann:2017gkg,Caloni:2024olo}. Overall, DESI-$P_{\rm 1D}$ tightens the constraint on $N_{\rm eff}$ by $3\%$--$7.5\%$, providing complementary information on the matter power spectrum. Although our compressed likelihood does not fully capture the scale dependence induced by free-streaming species, it remains sensitive to their effects on the amplitude and slope of the matter power spectrum. As a result, DESI-$P_{\rm 1D}$ consistently tightens the upper bound on $N_{\rm eff}$ and complements the BAO measurements, partly offsetting the weaker constraint obtained when BAO is included. This behavior agrees with the results reported in Ref.~\cite{Chaves-Montero:2026hqd}.

Other parameters, such as $n_s$, $S_8$, $\Delta_\star^2$, and $n_\star$, also benefit from the inclusion of DESI-P$_{\rm 1D}$. For example, allowing $N_{\rm eff}$ to vary introduces a degeneracy between the radiation density and the primordial spectrum, weakening the CMB-only constraint on $n_s$~\cite{Chaves-Montero:2026hqd}. DESI-P$_{\rm 1D}$ breaks this degeneracy, improving the precision on $n_s$. Similar gains occur for $\Delta_\star^2$ and $n_\star$. For $S_8$, the improvement comes from additional information on the matter fluctuation amplitude, though its reach is limited by the $z\simeq3$ redshift coverage.

Remarkably, the addition of DESI-P$_{\rm 1D}$ does not significantly shift the marginalized central values in Table~\ref{tab:Neffconstraints}. This supports our choice of dataset combination, as the main Lyman-$\alpha$ dataset is fully consistent with the CMB and BAO data, with no statistically significant tension between them. In fact, the consistency between the datasets can already be anticipated from the compressed parameters entering the likelihood. Using CMB data alone, we obtain
\begin{equation*}
  \Delta_\star^2=0.3552^{+0.0075}_{-0.0098},
  \qquad
  n_\star=-2.2958^{+0.0042}_{-0.0063},
\end{equation*}
whereas the DESI DR1 Lyman-$\alpha$ analysis~\cite{Chaves-Montero:2026hqd} finds
\begin{equation*}
  \Delta_\star^2=0.379\pm0.032,
  \qquad
  n_\star=-2.309\pm0.019.
\end{equation*}
The two determinations are statistically consistent. Estimating the differences using the marginalized uncertainties, we find a discrepancy of only about $0.7\sigma$. Thus, the DESI-P$_{\rm 1D}$ likelihood does not shift the CMB solution toward a substantially different region of parameter space. We find the same behavior for the CMB+BAO combination.

\begin{table*}[htpb!]
\centering
\renewcommand{\arraystretch}{1.4}
\resizebox{\textwidth}{!}{
\begin{tabular}{|l|c|c|c|c|}
\hline
\textbf{Parameter} &
\textbf{CMB} &
\textbf{CMB+DESI-P$_{\rm 1D}$} &
\textbf{CMB+DESI-BAO} &
\textbf{CMB+DESI-BAO+DESI-P$_{\rm 1D}$} \\
\hline

$\Omega_{\rm b} h^2$
& $0.02246^{+0.00016}_{-0.00018}$
& $0.02245\pm0.00016$
& $0.02265\pm0.00014$
& $0.02265\pm0.00014$ \\ 

$\Omega_{\rm c} h^2$
& $0.1216^{+0.0015}_{-0.0021}$
& $0.1217^{+0.0015}_{-0.0020}$
& $0.1205^{+0.0014}_{-0.0027}$
& $0.1203^{+0.0014}_{-0.0024}$ \\

$100\theta_s$
& $1.04161^{+0.00039}_{-0.00033}$
& $1.04159^{+0.00039}_{-0.00032}$
& $1.04168^{+0.00041}_{-0.00036}$
& $1.04171^{+0.00041}_{-0.00035}$ \\

$\tau_{\rm reio}$
& $0.0546\pm0.0075$
& $0.0553\pm0.0074$
& $0.0609\pm0.0073$
& $0.0617\pm0.0073$ \\

$\log(10^{10}A_{\rm s})$
& $3.102\pm0.014$
& $3.104\pm0.014$
& $3.101\pm0.015$
& $3.104\pm0.015$ \\

$n_s$
& $0.9690^{+0.0049}_{-0.0061}$
& $0.9687^{+0.0046}_{-0.0056}$
& $0.9760^{+0.0043}_{-0.0049}$
& $0.9755^{+0.0037}_{-0.0048}$ \\

$H_0$ [km/s/Mpc]
& $68.18^{+0.69}_{-0.99}$
& $68.14^{+0.68}_{-0.96}$
& $69.47^{+0.47}_{-0.81}$
& $69.43^{+0.47}_{-0.74}$ \\

$\Omega_m$
& $0.3114\pm0.0080$
& $0.3120\pm0.0077$
& $0.2979\pm0.0027$
& $0.2979\pm0.0027$ \\

$S_8$ 
& $0.831\pm 0.013$
& $0.832\pm 0.012$
& $0.8129\pm 0.0086$
& $0.8132\pm 0.0081$ \\

$\Delta_\star^2$
& $0.3552^{+0.0075}_{-0.0098}$
& $0.3561^{+0.0071}_{-0.0092}$
& $0.3544^{+0.0086}_{-0.012}$
& $0.3542^{+0.0080}_{-0.011}$ \\

$n_\star$
& $-2.2958^{+0.0042}_{-0.0063}$
& $-2.2960^{+0.0041}_{-0.0057}$
& $-2.2909^{+0.0048}_{-0.0061}$
& $-2.2915^{+0.0041}_{-0.0059}$ \\

$N_{\rm eff}$
& $<_{95\%} 3.36$
& $<_{95\%} 3.35$
& $<_{95\%} 3.44$
& $<_{95\%} 3.41$ \\

\hline
$\Delta \chi_{\rm min}^2$ & 0.84 & 0.80 & -1.34 & 0.60 \\
$\Delta \text{AIC}$ & 2.84 & 2.80 & 0.66 & 2.60 \\
$\ln \mathcal{B}_{ij}$ & 2.82 & 3.03 & 2.25 & 1.99 \\
\hline
\end{tabular}
}
\caption{Marginalized mean values with $68\%$ confidence intervals for each dataset combination, except for $N_{\rm eff}$, for which the $95\%$ CL upper limit is reported.}
\label{tab:Neffconstraints}
\end{table*}

In this context, DESI-BAO drives the largest shifts in the marginalized central values. Its geometric information complements that from the CMB and $P_{\rm 1D}$, helping to break parameter degeneracies. In particular, these shifts reflect the tension between the matter density preferred by DESI-BAO and that inferred from the CMB, with DESI-BAO favoring a lower value of $\Omega_m$~\cite{DESI:2025zgx,SPT-3G:2025bzu,Ferreira:2025lrd,Jhaveri:2025neg,Shlivko:2026jxa}. In the $\Lambda$CDM$+N_{\rm eff}$ model, it is well known that BAO data relax the $N_{\rm eff}$ constraint~\cite{Hou:2012xq,Allali:2024cji}. The lower $\Omega_m$ preferred by DESI-BAO is correlated with a higher $H_0$, while the relaxed upper limit on $N_{\rm eff}$ allows a larger early-time expansion rate and a smaller sound horizon. These correlated effects shift $H_0$ toward higher values, as observed in the CMB+DESI-BAO and CMB+DESI-BAO+DESI-$P_{\rm 1D}$ combinations. The direction of these shifts agrees with the degeneracy structure found in models with additional relativistic energy density~\cite{Planck:2018vyg,DiValentino:2024xsv,Escudero:2024uea}. We emphasize that, despite this increase, our tightest constraint from CMB+DESI-BAO+DESI-$P_{\rm 1D}$ remains in more than $4\sigma$ tension with the local measurement $H_0=73.50\pm0.81$ km/s/Mpc~\cite{H0DN:2025lyy}.

\begin{figure}[htpb!]
    \centering
    \includegraphics[width=0.5\textwidth]{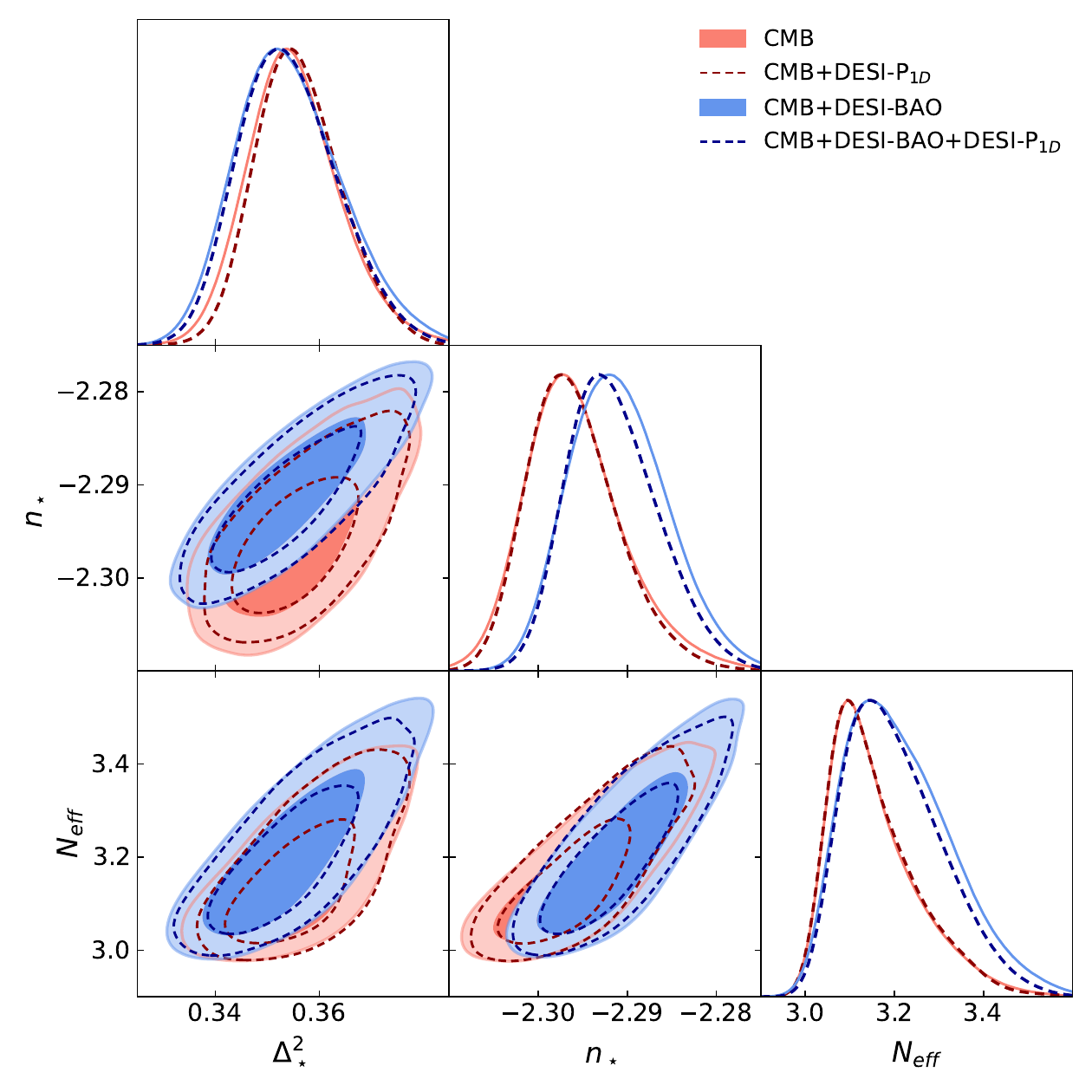}
    \includegraphics[width=0.49\textwidth]{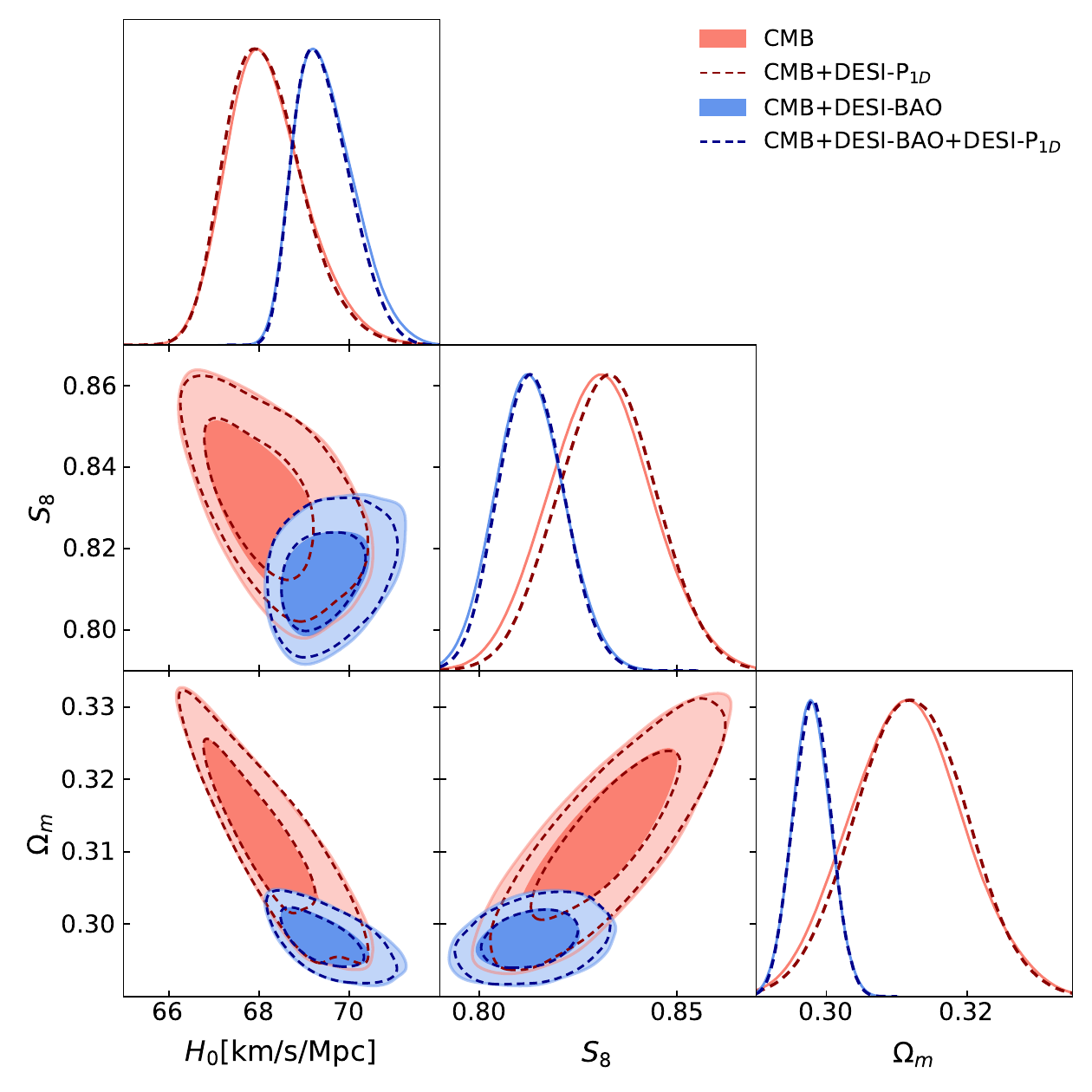}
    \caption{\textit{Left panel:} Marginalized one-dimensional posterior distributions and two-dimensional credible regions ($68\%$ and $95\%$) for the parameters $\Delta_\star^2$, $n_\star$, and $N_{\rm eff}$ within the $\Lambda$CDM+$N_{\rm eff}$ framework, using CMB data and its combinations with DESI-BAO and DESI-$P_{\rm 1D}$, as shown in the legend. \textit{Right panel:} Same as the left panel, but for the parameters $H_0$, $S_8$, and $\Omega_m$.}
    \label{fig:Neffconstraints}
\end{figure}

These shifts are also reflected in the other cosmological parameters. The posterior distributions in Fig.~\ref{fig:Neffconstraints} show these correlations. In particular, $H_0$ and $\Omega_m$ are anticorrelated, so the increase in $H_0$ is accompanied by lower values of $\Omega_m$. A similar trend appears in $\Lambda$CDM~\cite{DESI:2025zgx}, although it is weaker, with $\Omega_m$ shifting from $0.3169\pm0.0065$ to $0.3027\pm0.0036$. Allowing $N_{\rm eff}$ to vary slightly strengthens this effect. The additional freedom in the radiation density modifies the matter-radiation equality condition, allowing a further correlated adjustment of $\Omega_m$. This results in a difference of approximately $1.6\sigma$ between the $\Omega_m$ constraints from CMB and CMB+DESI-BAO+DESI-$P_{\rm 1D}$.

The lower value of $\Omega_m$ is also reflected in $S_8$, defined as $S_8=\sigma_8\sqrt{\Omega_m/0.3}$. As shown in Fig.~\ref{fig:Neffconstraints}, $S_8$ and $\Omega_m$ are positively correlated. Thus, the lower values of $\Omega_m$ preferred by the BAO data are accompanied by lower values of $S_8$. For the CMB+DESI-BAO+DESI-$P_{\rm 1D}$ combination, which gives our tightest constraint, we find $S_8=0.8132\pm0.0081$. Interestingly, this shift is in the direction of the lower $S_8$ values inferred from weak-lensing surveys. However, such a comparison should be interpreted with caution, since $S_8$ is a model-dependent derived parameter and weak-lensing constraints are generally obtained assuming a specific cosmological model~\cite{DES:2018ufa,KiDS:2020ghu,DES:2022ccp,Wright:2025xka,DES:2026fyc}. 

\smallskip

In summary, the constraints benefit from the inclusion of DESI-$P_{\rm 1D}$. The correlations in Fig.~\ref{fig:Neffconstraints} are consistent with those reported in the literature, including the positive $N_{\rm eff}$-$\Delta_\star^2$ and $N_{\rm eff}$-$n_\star$ correlations found in Ref.~\cite{Chaves-Montero:2026hqd}. Our constraints on $N_{\rm eff}$ are also consistent with recent studies~\cite{Giovanetti:2024eff,AtacamaCosmologyTelescope:2025nti,SPT-3G:2025bzu,Goldstein:2026iuu,Escudero:2026mgw,Yeh:2026pil}. For the model comparison, the $\Delta\chi^2_{\rm min}$ and $\Delta{\rm AIC}$ values show no significant preference for the extended model. The largest Bayesian preference for $\Lambda$CDM is obtained for CMB+DESI-$P_{\rm 1D}$, with $\ln\mathcal{B}_{ij}=3.03$, corresponding to strong evidence according to the scale adopted in Sec.~\ref{sec:statistical_tools}. The extended model improves the best fit only for CMB+DESI-BAO, with $\Delta\chi^2_{\rm min}=-1.34$. This modest improvement does not compensate for the additional parameter according to the AIC, for which $\Delta{\rm AIC}=0.66$, while the corresponding Bayes factor, $\ln\mathcal{B}_{ij}=2.25$, favors $\Lambda$CDM.

\subsection{Constraints on the DTS model}
\label{sec:DTS}

Our results for the DTS scenario are summarized in Table~\ref{tab:DTSconstraints}. We give marginalized means and $68\%$ credible intervals for all parameters, except for the neutrino-sector parameters $\Delta N_{\rm eff}$ and $m_s^{\rm eff}$, for which we report $95\%$ upper limits. In general, we find no significant evidence for a nonzero sterile-neutrino population. To better interpret this result, we use the relation in Eq.~\ref{eq:dts_delta_neff} to constrain the temperature ratio between sterile and active neutrinos. We obtain
\begin{equation}
\frac{T_s}{T_\nu} <
\begin{cases}
0.740, & {\rm CMB},\\
0.658, & {\rm CMB+DESI}\text{-}P_{\rm 1D},\\
0.697, & {\rm CMB+DESI\text{-}BAO},\\
0.753, & {\rm CMB+DESI\text{-}BAO+DESI}\text{-}P_{\rm 1D}.
\end{cases}
\end{equation}

All dataset combinations therefore constrain the temperature of a possible sterile-neutrino population to be lower than that of the fully thermalized case, $T_s/T_\nu=1$. Thus, if a thermal sterile-neutrino population is present, the data require its abundance to be suppressed relative to the fully thermalized case, consistent with previous constraints on partially populated or colder sterile-neutrino relics~\cite{Acero:2008rh,Das:2021pof,Ladeira:2026pck}.

\begin{table*}[htpb!]
\centering
\renewcommand{\arraystretch}{1.4}
\resizebox{\textwidth}{!}{
\begin{tabular}{|l|c|c|c|c|}
\hline
\textbf{Parameter} &
\textbf{CMB} &
\textbf{CMB+DESI-P$_{\rm 1D}$} &
\textbf{CMB+DESI-BAO} &
\textbf{CMB+DESI-BAO+DESI-P$_{\rm 1D}$} \\
\hline

$\Omega_{\rm b} h^2$
& $0.02241\pm0.00015$
& $0.02238\pm0.00015$
& $0.02261\pm0.00013$
& $0.02263\pm0.00013$ \\

$\Omega_{\rm c} h^2$
& $0.1212^{+0.0015}_{-0.0018}$
& $0.1209^{+0.0014}_{-0.0018}$
& $0.1176^{+0.0011}_{-0.0017}$
& $0.11902^{+0.00080}_{-0.0022}$ \\

$100\theta_s$
& $1.04165\pm0.00033$
& $1.04172^{+0.00034}_{-0.00029}$
& $1.04201^{+0.00034}_{-0.00028}$
& $1.04186^{+0.00040}_{-0.00031}$ \\

$\tau_{\rm reio}$
& $0.0548\pm0.0074$
& $0.0552\pm0.0074$
& $0.0620^{+0.0065}_{-0.0076}$
& $0.0637\pm0.0078$ \\

$\log(10^{10}A_{\rm s})$
& $3.107\pm0.015$
& $3.107\pm0.014$
& $3.104\pm0.015$
& $3.106\pm0.016$ \\

$n_s$
& $0.9660\pm0.0047$
& $0.9656\pm0.0043$
& $0.9726^{+0.0025}_{-0.0043}$
& $0.9746^{+0.0037}_{-0.0045}$ \\

$H_0$ [km/s/Mpc]
& $67.20\pm0.73$
& $67.30\pm0.59$
& $68.80^{+0.16}_{-0.45}$
& $69.08^{+0.30}_{-0.68}$ \\

$\Omega_m$
& $0.3229^{+0.0093}_{-0.013}$
& $0.3192^{+0.0078}_{-0.0088}$
& $0.2990\pm0.0025$
& $0.2987\pm0.0026$ \\

$S_8$ 
& $0.821^{+0.017}_{-0.013}$
& $0.829\pm 0.013$
& $0.799^{+0.013}_{-0.0088}$
& $0.8076\pm 0.0088$  \\

$\Delta_\star^2$
& $0.326^{+0.029}_{-0.011}$
& $0.341^{+0.013}_{-0.0089}$
& $0.332^{+0.022}_{-0.0092}$
& $0.3466^{+0.0084}_{-0.011}$ \\

$n_\star$
& $-2.307^{+0.013}_{-0.0050}$
& $-2.3035^{+0.0080}_{-0.0048}$
& $-2.304^{+0.015}_{-0.0040}$
& $-2.2939^{+0.0039}_{-0.0057}$ \\

$\Delta N_{\rm eff}$
& $<_{95\%} 0.30$
& $<_{95\%} 0.187$
& $<_{95\%} 0.236$ 
& $<_{95\%} 0.33$ \\

$m_s^{\rm eff}$
& $ <_{95\%} 0.41$ 
& $<_{95\%} 0.191$ 
& $<_{95\%} 0.335$ 
& $<_{95\%} 0.061$ \\

\hline

$\Delta \chi_{\rm min}^2$ & 1.98 & -0.18 & -2.16 & 0.18 \\
$\Delta \text{AIC}$ & 5.98 & 3.82 & 1.84 & 4.18 \\
$\ln \mathcal{B}_{ij}$ & 3.73 & 3.00 & 3.33 & 4.05 \\
\hline
\end{tabular}
}
\caption{Marginalized mean values with $68\%$ credible intervals for each dataset combination, except for $\Delta N_{\rm eff}$ and $m_s^{\rm eff}$, for which the $95\%$ credible upper limits are reported.}
\label{tab:DTSconstraints}
\end{table*}

Across the different dataset combinations, including DESI-P$_{\rm 1D}$ notably tightens the constraints on the DTS parameters. In the DTS scenario, DESI-P$_{\rm 1D}$ has a greater impact than in the $\Lambda$CDM+$N_{\rm eff}$ model (see Table~\ref{tab:Neffconstraints}), since it breaks additional degeneracies involving $m_s^{\rm eff}$. The full combination of CMB+DESI-BAO+DESI-P$_{\rm 1D}$ lowers the upper bound on $m_s^{\rm eff}$ by a factor of $\sim 7$ relative to CMB alone. Adding DESI-P$_{\rm 1D}$ to CMB alone improves the constraint by a factor of $\sim 2$. The full combination gives $m_s^{\rm eff}<0.061~{\rm eV}$.\footnote{Note that $m_s^{\rm eff}$ and $m_s$ are not the same quantity. They are related by Eq.~\ref{eq:dts_meff}, so a small $m_s^{\rm eff}$ does not necessarily indicate a small physical sterile-neutrino mass $m_s$.} Using Eq.~\ref{eq:dts_density}, this corresponds to
\begin{equation}
  \omega_s <_{95\%} 6.5\times10^{-4},
\end{equation}
which strongly limits the allowed late-time abundance of the massive sterile component.

Although the constraints on $m_s^{\rm eff}$ are striking, this parameter should not be considered in isolation because it is directly related to $\Delta N_{\rm eff}$ through Eq.~\ref{eq:dts_meff}. For the CMB+DESI-P$_{\rm 1D}$ combination, the upper bound on $\Delta N_{\rm eff}$ tightens by approximately $40\%$ compared with CMB alone. This gain is driven mainly by the additional P$_{\rm 1D}$ information, which helps break the degeneracy among $n_s$, $\Delta N_{\rm eff}$, and $m_s^{\rm eff}$. By contrast, adding DESI-P$_{\rm 1D}$ to CMB+DESI-BAO weakens the upper bound on $\Delta N_{\rm eff}$. A smaller $m_s^{\rm eff}$ together with a weaker constraint on $\Delta N_{\rm eff}$ may seem counterintuitive. However, this behavior is allowed by the correlation between the two parameters and their relation to the physical sterile-neutrino mass.

One possible explanation for the relaxed upper bound on $\Delta N_{\rm eff}$ is the different parameter degeneracies preferred by DESI-BAO and DESI-P$_{\rm 1D}$. Relative to CMB alone, adding DESI-P$_{\rm 1D}$ shifts $n_s$ slightly toward lower values, whereas adding DESI-BAO shifts it toward higher values. When CMB, DESI-BAO, and DESI-P$_{\rm 1D}$ are combined, the interplay between these different degeneracy directions broadens the allowed parameter space, including that of $\Delta N_{\rm eff}$.

\begin{figure}[htpb!]
    \centering
    \includegraphics[width=0.5\textwidth]{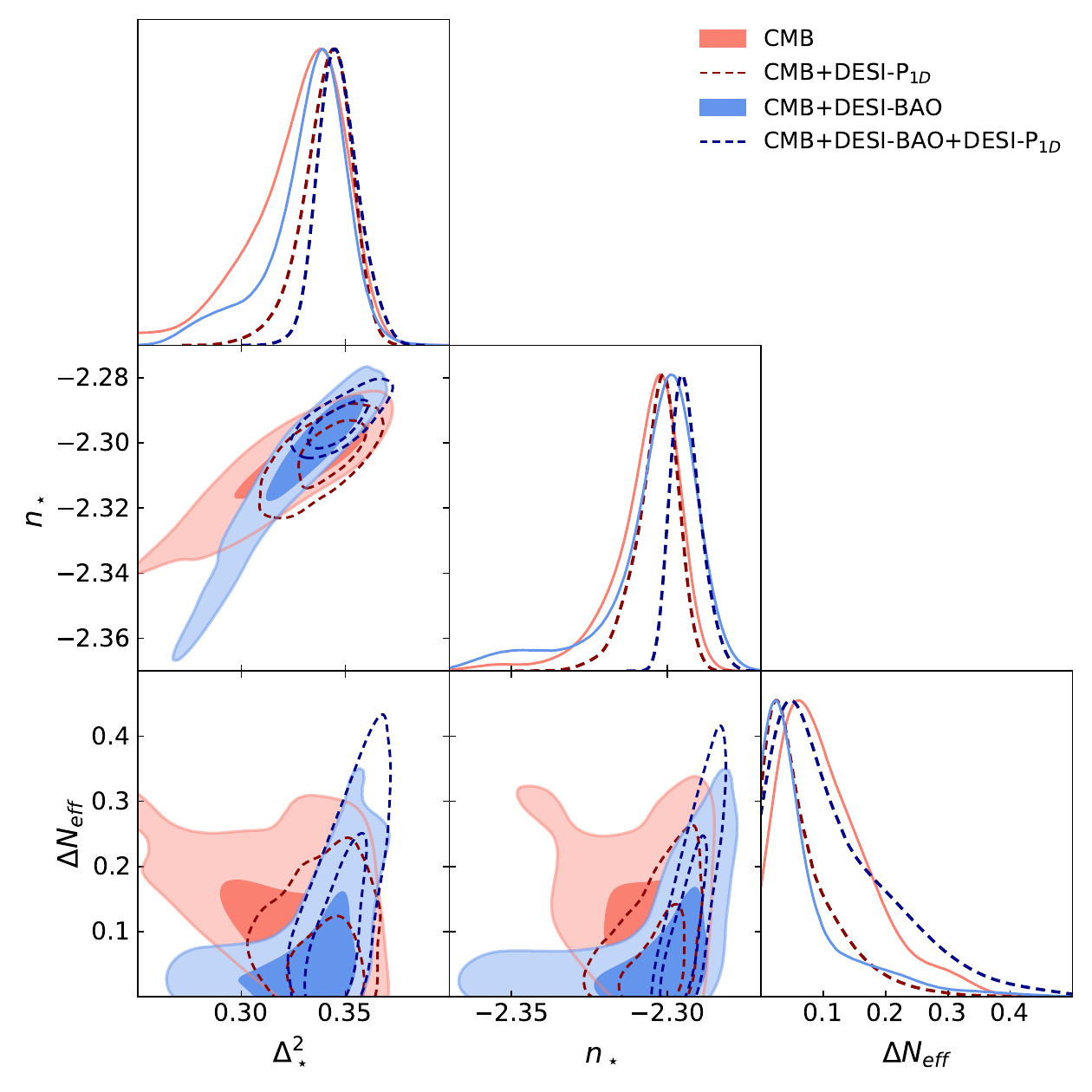}
    \includegraphics[width=0.49\textwidth]{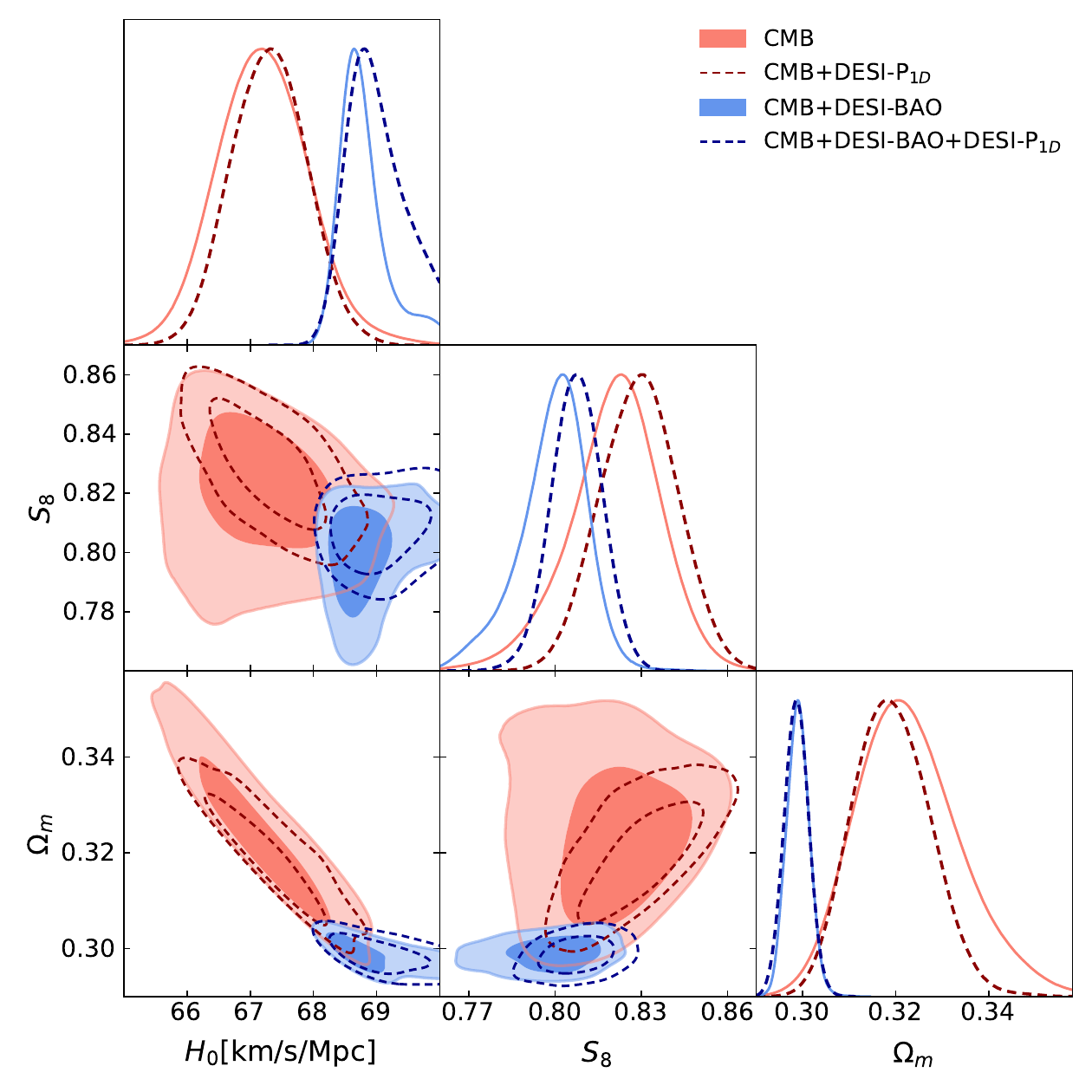}
    \caption{\textit{Left panel:} Marginalized one-dimensional posterior distributions and two-dimensional credible regions ($68\%$ and $95\%$) for the parameters $\Delta_\star^2$, $n_\star$, and $\Delta N_{\rm eff}$ within the DTS framework, using CMB data and its combinations with DESI-BAO and DESI-$P_{\rm 1D}$, as shown in the legend. \textit{Right panel:} Same as the left panel, but for the parameters $H_0$, $S_8$, and $\Omega_m$.}
    \label{fig:DTSconstraints}
\end{figure}

Figure~\ref{fig:DTSconstraints} shows the correlations among the model parameters. Relative to $\Lambda$CDM+$N_{\rm eff}$, the posterior distributions exhibit stronger degeneracies, as expected given the additional degree of freedom. The background parameters, however, show similar central values and correlations to those in the $\Lambda$CDM+$N_{\rm eff}$ model (discussed in Sec.~\ref{sec:Neff}). As before, adding BAO data shifts the inferred value of $H_0$ toward higher values, although it remains below the local measurement. The change in $S_8$ is more pronounced in the DTS scenario: including BAO data lowers it to $S_8=0.799^{+0.013}_{-0.0088}$ for the CMB+DESI-BAO combination. Interestingly, this shift is in the direction of the lower $S_8$ values inferred from weak-lensing surveys, such as KiDS-Legacy and DES Y6 $3\times2$pt~\cite{Wright:2025xka,DES:2026fyc}. However, such a comparison should be interpreted with caution, since $S_8$ is a model-dependent derived parameter and these weak-lensing constraints are obtained assuming a specific cosmological model.

Overall, our findings for the neutrino sector in the DTS scenario agree with recent studies~\cite{Pan:2023frx,Du:2025iow,Ladeira:2026pck}. Adding DESI-P$_{\rm 1D}$ strengthens the constraints, particularly on $m_s^{\rm eff}$. For the CMB+DESI-BAO+DESI-P$_{\rm 1D}$ combination, we obtain $m_s^{\rm eff}<0.061~{\rm eV}$, among the most restrictive upper bounds on this model reported in the recent literature. Because of this tight limit, the background parameters remain largely unchanged from those in $\Lambda$CDM+$N_{\rm eff}$. Our results also constrain the temperature of a possible sterile-neutrino population to be lower than in the fully thermalized case. Concerning model comparison, although the DTS model provides a slightly better fit than $\Lambda$CDM for CMB+DESI-P$_{\rm 1D}$ and CMB+DESI-BAO, $\Lambda$CDM is preferred once model complexity is accounted for. This preference is particularly clear from the Bayes factor $\ln\mathcal{B}_{ij}$, which favors $\Lambda$CDM in all cases.\\

\subsection{Thermal-relic interpretation of the $N_{\rm eff}$ constraint}
\label{sec:thermal_interpretation}

We translate the upper limits on $\Delta N_{\rm eff}$ into constraints on the thermal history of possible light relics. Following the thermal-relic treatment adopted by Planck~\cite{Planck:2018vyg}, we relate the upper limits on $\Delta N_{\rm eff}$ to the entropy degrees of freedom at decoupling and use the Standard Model equation of state of Ref.~\cite{Borsanyi:2016ksw} to obtain the corresponding lower limits on $T_{\rm d}$.

After decoupling, $T_X\propto a^{-1}$, while the species that remain in thermal equilibrium share the entropy released as the Standard Model plasma cools. The entropy density can be written as

\begin{equation}
s_{\rm SM}(T)
=
\frac{2\pi^2}{45}
g_{*s}^{\rm SM}(T)T^3 ,
\label{eq:entropy_density}
\end{equation}\smallskip
and conservation of comoving entropy gives~\cite{Planck:2018vyg,Dvorkin:2022jyg}

\begin{equation}
\frac{T_X}{T_\nu}
=
\left[
\frac{g_{*s}^{\rm SM}(T_{\nu,\rm dec})}
     {g_{*s}^{\rm SM}(T_{\rm d})}
\right]^{1/3}
\simeq
\left[
\frac{43}
     {4g_{*s}^{\rm SM}(T_{\rm d})}
\right]^{1/3}.
\label{eq:thermal_temperature_ratio}
\end{equation}\smallskip

The last expression uses $g_{*s}^{\rm SM}(T_{\nu,\rm dec})=43/4$, as obtained in the instantaneous neutrino-decoupling approximation. The contribution of $X$ to the radiation density is then

\begin{equation}
\Delta N_{\rm eff}
=
g_X C_X
\left(\frac{T_X}{T_\nu}\right)^4
=
g_X
\left[
\frac{43}{4g_{*s}^{\rm SM}(T_{\rm d})}
\right]^{4/3}
C_X,
\qquad
C_X=
\begin{cases}
4/7, & {\rm boson},\\
1/2, & {\rm fermion},
\end{cases}
\label{eq:thermal_delta_neff}
\end{equation}
where $g_X$ is the number of internal degrees of freedom of the relic and $T_{\rm d}$ is its decoupling temperature. This is the relation used in the Planck thermal-relic analysis~\cite{Planck:2018vyg}. Inverting Eq.~\eqref{eq:thermal_delta_neff} gives

\begin{equation}
g_{*s}^{\rm SM}(T_{\rm d})
=
\frac{43}{4}
\left[
\frac{g_X C_X}{\Delta N_{\rm eff}}
\right]^{3/4}.
\label{eq:gst_inversion}
\end{equation}\smallskip

In the high-temperature ideal-gas limit, $g_{*s}^{\rm SM}=106.75$, the commonly quoted minimum thermal contributions are~\cite{Baumann:2017gkg,Dvorkin:2022jyg,Euclid:2024imf}

\begin{equation}
\Delta N_{\rm eff}^{\rm min}
\simeq
\begin{cases}
0.027, & {\rm real\ scalar},\\
0.047, & {\rm Weyl\ fermion},\\
0.054, & {\rm massless\ gauge\ boson}.
\end{cases}
\label{eq:thermal_benchmarks}
\end{equation}\smallskip

We take $g_X=1$ for a real scalar and $g_X=2$ for both a Weyl fermion and a massless gauge boson. These three cases provide simple benchmarks for different classes of thermal relics. The scalar case can describe an axion- or Goldstone-like degree of freedom, the Weyl case a light chiral fermion such as a right-handed neutrino, and the vector case a massless dark photon~\cite{Dvorkin:2022jyg,Euclid:2024imf}.

For each dataset combination, we evaluate Eq.~\eqref{eq:gst_inversion} at the 95\% upper limit $\Delta N_{\rm eff}^{95}$,

\begin{equation}
g_{*s}^{\rm min}
=
\frac{43}{4}
\left[
\frac{g_X C_X}{\Delta N_{\rm eff}^{95}}
\right]^{3/4},
\label{eq:gst_lower_bound}
\end{equation}
which implies
$g_{*s}^{\rm SM}(T_{\rm d})>g_{*s}^{\rm min}$.
The corresponding minimum decoupling temperature satisfies

\begin{equation}
g_{*s}^{\rm SM}(T_{\rm d}^{\rm min})
=
g_{*s}^{\rm min}.
\label{eq:td_from_gstar}
\end{equation}

To evaluate Eq.~\eqref{eq:td_from_gstar}, we use the Standard Model equation of state of Ref.~\cite{Borsanyi:2016ksw}, following the treatment adopted in the Planck analysis~\cite{Planck:2018vyg}. Borsanyi et al. provide a tabulation in $\log_{10}(T/{\rm MeV})$ of $g_\rho(T)$ and $g_\rho(T)/g_{*s}(T)$. We interpolate these quantities with a cubic spline and compute

\begin{equation}
g_{*s}^{\rm SM}(T)
=
\frac{g_\rho(T)}
     {g_\rho(T)/g_{*s}(T)}.
\label{eq:gs_borsanyi}
\end{equation}
We then solve Eq.~\eqref{eq:td_from_gstar} numerically. Since the conversion is obtained from an interpolated equation of state, the temperatures below are quoted as approximate bounds.

For example, the full CMB+DESI-BAO+DESI-$P_{\rm 1D}$ combination gives $\Delta N_{\rm eff}^{95}=0.366$. For a real scalar, $g_XC_X=4/7$, which gives $g_{*s}^{\rm min}=15.01$. The Standard Model equation of state then gives $T_{\rm d}^{\rm min}\simeq62$ MeV. The same procedure is applied to the other benchmarks and dataset combinations, with the resulting bounds summarized in Table~\ref{tab:thermal_interpretation}.

For a fixed relic, decoupling at a lower temperature corresponds to a smaller $g_{*s}^{\rm SM}(T_{\rm d})$ and therefore to a larger $\Delta N_{\rm eff}$. The upper limits on $\Delta N_{\rm eff}$ thus set a minimum decoupling temperature. For the full data combination, these limits are $T_{\rm d}\gtrsim62$ MeV, $141$ MeV, and $150$ MeV for the real-scalar, Weyl-fermion, and massless-vector benchmarks, respectively.

The bounds reach the temperature range of the QCD crossover, where the entropy degrees of freedom of the Standard Model plasma change rapidly~\cite{Borsanyi:2016ksw}. For the full combination, the Weyl-fermion and massless-vector limits lie close to the pseudocritical temperature, $T_c=(156.5\pm1.5)$ MeV~\cite{HotQCD:2018pds}, while the real-scalar case allows decoupling at lower temperatures. \textit{A relic that decouples at a higher temperature does not share the subsequent entropy release of the plasma and therefore gives a smaller contribution to $\Delta N_{\rm eff}$.}

The values in Eq.~\eqref{eq:thermal_benchmarks} remain below all the 95\% upper limits obtained here. The three benchmark relics can therefore satisfy the cosmological bounds if they decouple sufficiently early. \textit{Since $\Delta N_{\rm eff}=0$ also remains allowed, these results do not constitute evidence for an additional light relic or a measurement of $T_{\rm d}$. They give lower limits on the decoupling temperature for each thermal benchmark under the assumptions stated above.}

\begin{table*}[htpb!]
\centering
\renewcommand{\arraystretch}{1.4}
\resizebox{\textwidth}{!}{
\begin{tabular}{|l|c|cc|cc|cc|}
\hline
Datasets
& $\Delta N_{\rm eff}$ (95\%)
& \multicolumn{2}{c|}{Real scalar}
& \multicolumn{2}{c|}{Weyl fermion}
& \multicolumn{2}{c|}{Massless gauge boson}
\\
\cline{3-8}
&
& $g_{*s}^{\rm min}$ & $T_{\rm d}^{\rm min}$ [MeV]
& $g_{*s}^{\rm min}$ & $T_{\rm d}^{\rm min}$ [MeV]
& $g_{*s}^{\rm min}$ & $T_{\rm d}^{\rm min}$ [MeV]
\\
\hline

CMB
& $<0.316$
& $>16.76$ & $\gtrsim92$
& $>25.51$ & $\gtrsim151$
& $>28.19$ & $\gtrsim160$
\\

CMB+DESI-P$_{\rm 1D}$
& $<0.306$
& $>17.17$ & $\gtrsim99$
& $>26.13$ & $\gtrsim153$
& $>28.88$ & $\gtrsim163$
\\

CMB+DESI-BAO
& $<0.396$
& $>14.15$ & $\gtrsim49$
& $>21.53$ & $\gtrsim136$
& $>23.80$ & $\gtrsim145$
\\

CMB+DESI-BAO+DESI-P$_{\rm 1D}$
& $<0.366$
& $>15.01$ & $\gtrsim62$
& $>22.85$ & $\gtrsim141$
& $>25.25$ & $\gtrsim150$
\\

\hline
\end{tabular}}
\caption{
Thermal-relic interpretation of the 95\% upper limits on $\Delta N_{\rm eff}$. The lower limits on $g_{*s}^{\rm SM}(T_{\rm d})$ follow from Eq.~\eqref{eq:gst_lower_bound}. The corresponding limits on $T_{\rm d}$ are obtained using the Standard Model equation of state of Ref.~\cite{Borsanyi:2016ksw}, following the thermal-relic treatment of Ref.~\cite{Planck:2018vyg}. The calculation assumes thermal equilibrium with the Standard Model followed by relativistic decoupling.
}
\label{tab:thermal_interpretation}
\end{table*}

\section{Final Remarks}
\label{sec:conclusions}

In this work, we investigated constraints on two extensions of the standard neutrino sector, the $\Lambda$CDM$+N_\mathrm{eff}$ model and the DTS framework, using the DESI DR1 one-dimensional Lyman-$\alpha$ forest power spectrum, combined with Planck 2018 CMB and DESI DR2 BAO measurements. Before performing the cosmological inference, we validated the two-parameter compression scheme ($\Delta^2_\star$, $n_\star$) for the DTS model, finding that the residual cosmological dependence after rescaling remains below $0.15\%$, confirming its applicability to this class of non-standard scenarios. For both models, no significant evidence for additional radiation or a sterile component was found; all constraints are consistent with the Standard Model prediction $N_\mathrm{eff}^\mathrm{SM} = 3.044$ and with a cosmologically negligible sterile abundance. The Lyman-$\alpha$ data consistently tighten the constraints from CMB and BAO alone, mainly by breaking degeneracies involving $n_s$, $N_\mathrm{eff}$, and $m_s^\mathrm{eff}$, while the compressed parameters inferred from CMB and DESI DR1 Lyman-$\alpha$ are statistically consistent at the $\lesssim 0.7\sigma$ level. Model-comparison statistics show no evidence in favor of either extension, with the AIC and Bayes factors generally favoring $\Lambda$CDM. The upper limits on $\Delta N_\mathrm{eff}$ were further interpreted in terms of thermal light relics.

Looking ahead, the constraints presented here will be substantially improved by upcoming surveys across both CMB and large-scale structure (LSS) observables. On the CMB side, the Simons Observatory~\cite{Banerjee:2025gwe} and CMB-S4~\cite{CMB-S4:2026mge} are expected to reach $\sigma(N_\mathrm{eff}) \simeq 0.05$ and $\sigma(N_\mathrm{eff}) \simeq 0.02$--$0.03$, respectively.

For the DTS model, the gain in $\sigma(N_\mathrm{eff})$ will translate into improved constraints on $\Delta N_\mathrm{eff}^s$ and, through its correlation with the sterile-neutrino abundance, on $m_s^\mathrm{eff}$, potentially pushing the bound below the minimum active-neutrino mass scale. Higher-resolution CMB experiments will additionally provide more precise lensing reconstructions, which are sensitive to the suppression of small-scale structure induced by both active- and sterile-neutrino free-streaming.

On the LSS side, several complementary probes will extend the reach of the present analysis. Full-shape analyses of the galaxy power spectrum from DESI, combining information on the BAO scale, the growth rate, and the broadband shape of $P(k)$, are sensitive to the epoch of matter-radiation equality and therefore to $N_\mathrm{eff}$ through the characteristic shift it induces in the turnover scale~\cite{Baumann:2017gkg,Elbers:2025vlz}. The combination of DESI full-shape with CMB and Lyman-$\alpha$ is expected to improve the $N_\mathrm{eff}$ constraints. The Euclid satellite~\cite{Euclid:2019clj} will provide galaxy clustering and weak-lensing measurements over a wide redshift range with exquisite statistical precision. Its projected sensitivity $\sigma(\sum m_\nu) \sim 0.01$--$0.02~\mathrm{eV}$ makes it particularly powerful for constraining the DTS scenario, where the effect of $m_s^\mathrm{eff}$ on $P(k)$ is analogous to that of the active-neutrino mass sum but with a free-streaming scale determined by the sterile-neutrino mass and temperature. A joint analysis of Euclid clustering and lensing with CMB-S4 will help break the $m_s$--$T_s$ degeneracy that cannot be resolved by current data. The Rubin Observatory LSST will independently contribute through weak gravitational lensing and photometric galaxy clustering over a large sky area, providing a powerful cross-check on the matter power spectrum shape at low and intermediate redshifts.

The Lyman-$\alpha$ forest itself will see major improvements. Future DESI data releases will reduce the statistical uncertainties of the one-dimensional power spectrum relative to DR1. At that level, the systematic budget, in particular the modeling of the intergalactic medium temperature, pressure, and ionization, will become increasingly important. Extensions of the compression scheme beyond the current two-parameter description, or the use of simulation-based inference with emulators trained on wider parameter spaces, may be needed to fully exploit this information for non-standard neutrino scenarios such as DTS.

Together, these next-generation datasets will transform the landscape of neutrino-sector constraints. The multi-probe approach combining CMB polarization, galaxy full-shape, weak lensing, and the Lyman-$\alpha$ forest will either reveal signatures of light thermal relics or place increasingly stringent limits on their abundance, probing the thermal history of the early Universe.

\acknowledgments
E.S. received support from the CAPES scholarship. R.C.N. acknowledges financial support from the Conselho Nacional de Desenvolvimento Científico e Tecnológico (CNPq, National Council for Scientific and Technological Development) through Grant No. 304306/2022-3, and partial financial support from the Fundação de Amparo à Pesquisa do Estado do Rio Grande do Sul (FAPERGS, Research Support Foundation of the State of Rio Grande do Sul) through Grant No. 23/2551-0000848-3 and Grant No. 25/2551-0002612-1.
E.D.V. is supported by a Royal Society Dorothy Hodgkin Research Fellowship.
This article is based upon work from COST Action CA21136 \emph{Addressing observational tensions in cosmology with systematics and fundamental physics} (CosmoVerse), supported by COST (European Cooperation in Science and Technology).

\bibliographystyle{JHEP}
\bibliography{main}
\end{document}